# Energy limit for linear-to-nonlinear femtosecond laser pulse focusing in air

Yu.E.Geints[1], D.V.Mokrousova[2], D.V.Pushkarev[2], G.E.Rizaev[2],
L.V.Seleznev[2], I.Yu.Geints[1,3], A.A.Ionin[2] and A.A.Zemlyanov[1],

[1]*V. E. Zuev Institute of Atmospheric Optics of Siberian Branch of the Russian Academy of Sciences, 1, Academician Zuev Square, Tomsk 634055, Russia*
[2]*P. N. Lebedev Physical Institute of the Russian Academy of Sciences, 53 Leninskii pr., Moscow 119991, Russia*
[3]*Faculty of Physics, Lomonosov Moscow State University, Leninskie Gory, Moscow 119991, Russia*

\* e-mail: ygeints@iao.ru

**Abstract**
Propagation of atightly focused high-power ultrashort laser pulse in an optical medium is usually substantially influenced by the medium optical nonlinearity that can noticeably affect the laser pulse parameters around the nonlinear focus and lead to unavoidable and often undesirable spatial distortions of the focal waist. We present the results of our experimental study and numerical simulations on a femtosecond Ti:Sapphirelaser pulse propagation in air under different spatial focusing. We concentrated our study on spectral-angular and spatial pulse transformations under different focusing regimes - from linear to nonlinear one, when pulse filamentation occurs. For the first time to the best of our knowledge, we found the laser pulse numerical apertures range – namely, from $NA = 2 \cdot 10^{-3}$ to $5 \cdot 10^{-3}$ (for the laser pulse energy of 1 mJ), where the laser pulse distortions both in frequency-angular spectrum and pulse spatial shape are minimal. By means of the numerical simulations, we found the threshold pulse energy and peak power in a wide range of focusing conditions, within which a transition between the linear and strongly nonlinear laser pulse focusing in air takes place. This energy limit is shown to decrease with pulse numerical aperture enhancement. Our findings identify the laser pulse numerical apertures and energes adequate for getting a maximum laser intensity with a good beam quality around the focal point suitable for various laser micropatterning and micromachining technologies.

## 1. Introduction

Strong nonlinearity of an optical medium usually plays a noticeable role in a high peak power laser pulse propagation in this medium, which results in the pulse spatio-temporal self-modulation and large-scale changes of its spectral composition taking place in the areas of the pulse high intensity, namely, in the laser beam filaments accompanying by elongated plasma channels with relatively high free-electron density. In air and other transparent media like water, solid dielectrics, etc., the peak intensity in such a filament can reach up to hundreds of TW/cm$^2$, while an average filament lateral size varies from units to hundreds of micrometers depending on the propagation medium, laser wavelength and focusing conditions [1]. During the filamentation, a deep self-phase modulation of the laser pulse takes place, which results in significant enrichment of its frequency-angular spectra. This also leads to formation of wide supercontinuum wings [2] and high divergent rings of conical emission [3]. Up to now, a plenty of studies has been devoted to the filamentation of ultrashort laser pulses and its possible applications (see, for example, reviews [1, 4, 5]).

During propagation of a collimated or focused laser pulse with peak power $P_0$ exceeding the critical one for self-focusing $P_c$, the filamentation starts in the so-called nonlinear focus. The distance $z_{sf}$ to the nonlinear focus can be fairly accurately estimated using semi-empirical Marburger formula

[6]. This formula establishes the inverse proportional dependence of the self-focusing distance on the laser pulse power: $z_{sf} \propto 1/\sqrt{\eta}$, where the dimensionless parameter $\eta = P_0/P_c$ defines the reduced laser pulse power. Importantly, in the case of unstable operation of a femtosecond laser system, its peak power $P_0$ may change randomly within some range that inevitably leads to similar random displacements of the filamentation onset distance. This may be undesirable for application of ultrashort laser pulses in various laser technologies of materials processing [7] and surface micro-structuring [8], when the focal spot position must be known with high accuracy. That is why in such applications spatially pre-focused laser beams are applied in order to spatially stabilize the focal area.

As a matter of fact, according to the generalized Marburger formula the resulting focus position $f_N$ of a high-power laser beam focused by a linear concave mirror (or a lens) with the focusing distance $f$ after its propagation through a nonlinear medium can be found according to the reciprocal rule: $1/f_N = 1/z_{sf} + 1/f$ [9]. Thus, under the condition of relatively tight focusing $f \ll z_{sf}$, the resulting focus position is close to the linear one, i.e. $f_N \cong f$. However, a laser pulse extremely high intensity results in multiphoton/tunnel medium ionization in the geometrical focal spot and around it leading to a plasma cloud formation, which can substantially distort all the focusing geometry. In particular, F. Théberge et al. [10] demonstrated that under external linear focusing of a supercritical femtosecond pulse in air the increase of initial numerical aperture (*NA*) results in several orders of magnitude growth of laser plasma density in the nonlinear focal spot. Later Y. Geints et al. [11] experimentally and numerically demonstrated that not only plasma density, but also transverse and longitudinal dimensions of the nonlinear focal spot did change under tight focusing. In other words, the focal area and optical intensity significantly depend on the initial laser pulse power and laser beam *NA*. An analytical method of conditional classification for a laser pulse nonlinear focusing under the Kerr and plasma nonlinearities was suggested in Refs. [12, 13]. As was shown, in the transitional range of the laser pulse numerical apertures $NA = D_0/2f$ from $\sim 3 \cdot 10^{-3}$ to $5 \cdot 10^{-3}$ - where $D_0$ is the initial laser beam diameter determined by the fluence profile at the level of $1/e$ - partial contributions by the Kerr effect and plasma cloud to the optical wave phase become equal. This defines the boundary between so-called «linear» and «nonlinear» focusing regimes, which furthermore weakly depends on the laser pulse parameters such as its peak power and temporal duration.

In this paper, we are considering the filamentation of a tightly focused femtosecond laser pulse just by analyzing the dynamics of nonlinear focus formation and the evolution of the laser pulse parameters. By means of laboratory and numerical experiments, we study the laser pulse frequency-angular spectra and thoroughly consider the laser pulse spatial structure and distribution of the plasma cloud emerging due to multiphoton air ionization in the focal region at different focusing regimes. We founded that the transition boundary between different focusing regimes reported earlier in [12] actually corresponded to the lowest degree of nonlinear transformations of the focused laser pulse under its self-focusing including its frequency-angular spectrum and spatial dimensions of the focal spot. Our results can be interesting for some practical applications, when minimization of the focus position distortion caused by laser plasma is crucial. We demonstrate that for each *NA* the threshold pulse power and energy providing an insignificant degree of the plasma-driven distortions can be determined. These threshold energy are shown to decrease with the growth of the laser pulse *NA*.

## 2. Experiments

The experiments are carried out with the Ti:Sapphire laser system (Avesta Ltd.) producing pulses with the carrier wavelength of 744 nm, FWHM duration of 100 fs and repetition rate of 10 Hz. The laser beam $1/e$-diameter ($D_0$) at the laser system output is either 8 mm or 3 mm if telescoped. The *NA* of the laser pulse can be changed by using different lenses with the focal distances $f$ from

4 cm to 2.6 m. A linear distribution of the plasma density $\rho_{ez}$ produced in air along the pulse propagation axis is measured by the capacity-probe technique described in details in [14].

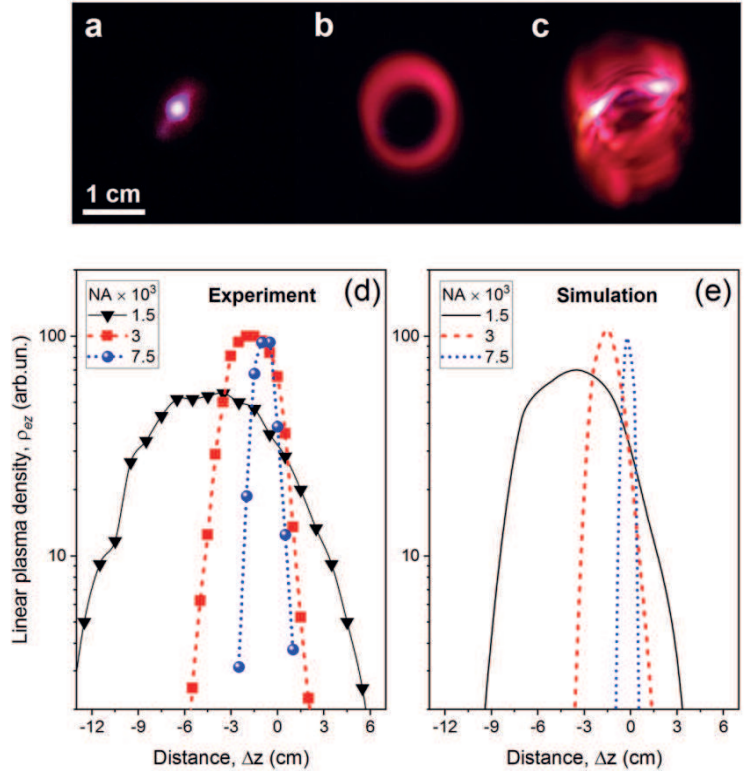

Figure 1.(a-c) Photographic images of 1 mJ laser pulse cross-sections on a paper screen located at the distance of four focal lengths $f$ corresponding to numerical apertures: $NA = 1.8 \cdot 10^{-3}$ (a), $8.8 \cdot 10^{-3}$ (b) and $21 \cdot 10^{-3}$ (c); (d, e). Linear plasma density $\rho_{ez}$ along the propagation axis versus longitudinal shift $\Delta z = (z-f)$ for different numerical apertures $NA$ ($E = 1$ mJ). The coordinate $\Delta z = 0$ corresponds to the linear focus.

In Figs. 1 (a)-(c) the photographic images of the laser pulse cross-sections obtained after focal filamentation on a paper screen in the far-field zone ($z = 4f$) under different focusing are presented. The images are taken with 3 mm diameter laser beam. The pulse energy $E$ is 1 mJ that corresponds to approximately three critical powers $P_c$ for self-focusing in air, if the value $P_c = 3.2$ GW is adopted [15]. While using relatively weakfocusing ($NA = 1.8 \cdot 10^{-3}$, Fig. 1(a)), a single intensity maximum in the optical transverse profileis visible corresponding to the post-filamentation channel [16]. Under the most tight focusing ($NA = 21 \cdot 10^{-3}$), one can see in Fig. 1(c) a complicated polychromatic laser pulse cross-section image formed by the field interference of conical emission rings generated around two individual filaments - the bright white spots - in the focal area. Note, that similar pulse imagecan be found elsewhere, e.g., in Ref. [17]. In the intermediate case of moderate focusing ($NA = 8.8 \cdot 10^{-3}$, Fig. 1(b)), the quasi-monochromatic ring of conical emission is clearly evident in the picture. This ring is predominantly of red color and arises due to the laser pulse self-phase modulation during the filamentation and subsequent broad-angle refraction on the on-axis plasma cloud. In this pulse image the darker central area is a consequence of low sensitivity of our digital camera in the infrared spectral range (> 700 nm) that indicates insignificant spectral transformations of the main axial pulse area during the filamentation under focusing.

Considering Figs. 1 (a)-(c) we can conclude that after the filamentation of spatially pre-focused laser beam, both its spectral composition and angular distribution strongly depend on $NA$. Therefore, our further study is directed towards analyzing the dependance of the laser pulse frequency-angular spectra on different focusing conditions.

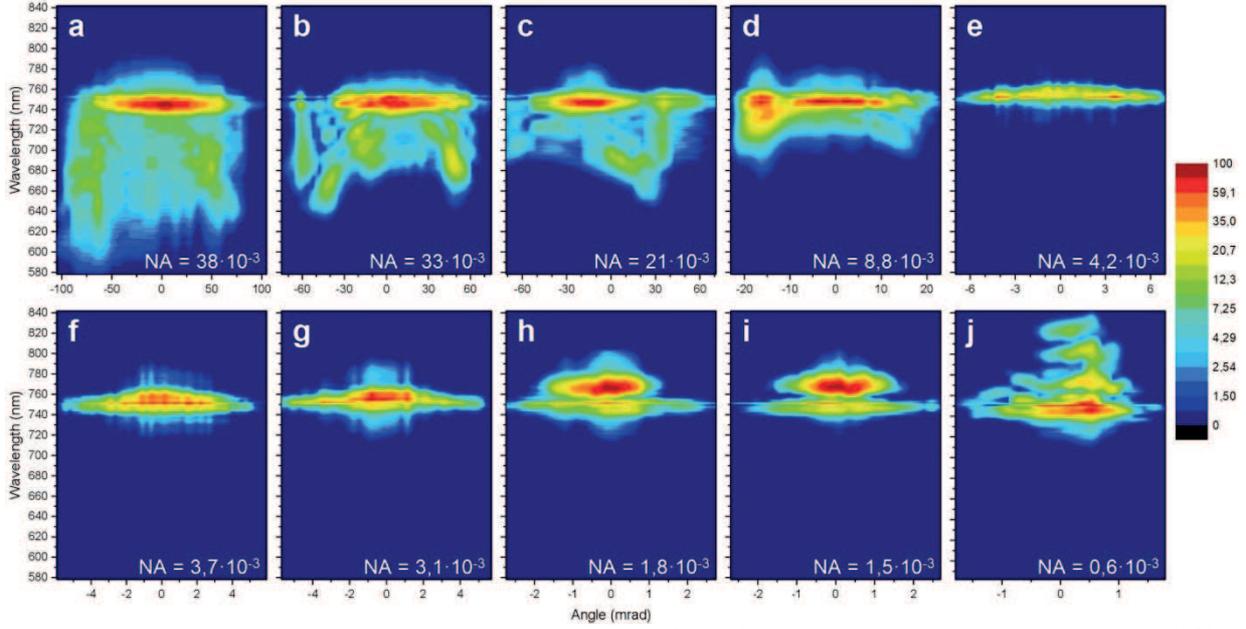

Рис.2. (a-j) Experimental frequency-angular spectra of the femtosecond pulse ($E = 1$ mJ) at different numerical apertures $NA$. The laser beam spatial diameter $D_0$ is 3 mm except for (b), (f) and (i), where $D_0$ is 8 mm.

We measure the frequency-angular spectra in the far-field at the distance of about four focal lengths. After the filamentation region normally located near the geometrical focus, a glass plate attenuates and directs the laser pulse to a spectrometer input slit. The spectrometer (Avesta-ASP150) is mounted on a metal rail and can be moved perpendicular to the direction of laser propagation, thus scanning step by step the cross-section of the laser beam in a horizontal plane. At each position of the spectrometer, only the light propagating at the given angle $\gamma$ to the optical axis enters the spectrometer slit. The deflection angle $\gamma$ is calculated as the ratio of the spectrometer perpendicular shift to the distance measured from the lens geometrical focus. For each $NA$ we obtain more than 30 experimental spectra taken at different spatial angles. By combining these spectra, we construct the resulting frequency-angular pulse spectrum for a given $NA$ as shown in Fig. 2.

These spectra can be conditionally divided into three groups. The first group presented in Figs. 2(a)–(d), contains the spectra recorded at high numerical apertures $NA > 5 \cdot 10^{-3}$ and corresponds to the pulse filamentation in the so-called linear focusing regime [12]. Under such conditions, during the laser pulse filamentation a plasma channel generated in the focal area is sufficiently dense with free-electrons concentration exceeding $10^{18}$ cm$^{-3}$ [10, 11]. However, the length of the plasma channel is relatively short as is seen by blue dotted line in Fig. 1(d), where the free-electron linear density $\rho_{ez}$ is plotted for different $NA$. Such high plasma density provides considerable pulse spectral broadening to the blue region. Besides, the back edge of the laser pulse is refracted on this dense plasma channel that causes spectrally broadened pulse propagation at large spatial angles $\gamma$ to the optical axis.

The laser pulse spectra shown in Figs. 2 (h)–(j) can be combined together as ones corresponding to weak spatial focusing with low numerical apertures $NA < 2 \cdot 10^{-3}$. In this group, the plasma areas are noticeably – by several times - longer, whereas the peak plasma density is significantly lower in comparison with the first spectrum group as is evidenced from the plasma distribution for $NA = 1.5 \cdot 10^{-3}$ in Fig. 1(d) by black solid line. The substantial elongation of the spatial region, where active pulse self-phase modulation takes place, in this case results in an increase of the impact of the retarded Kerr effect. Physically, this can be treated as the stimulated ro-vibrational Raman scattering of pulse spectral components during the propagation causing a pronounced red shift of pulse spectrum in the form of Raman humps [18]. Within the $NA$ range under consideration, the decrease of *NA* leads to the appearance of additional maxima in the long-wavelength region (Fig. 2 (h)), which was also observed earlier in [19]. At the same time, the lower laser plasma density in the focal area mitigates the pulse spectra broadening in the anti-Stokes wing.

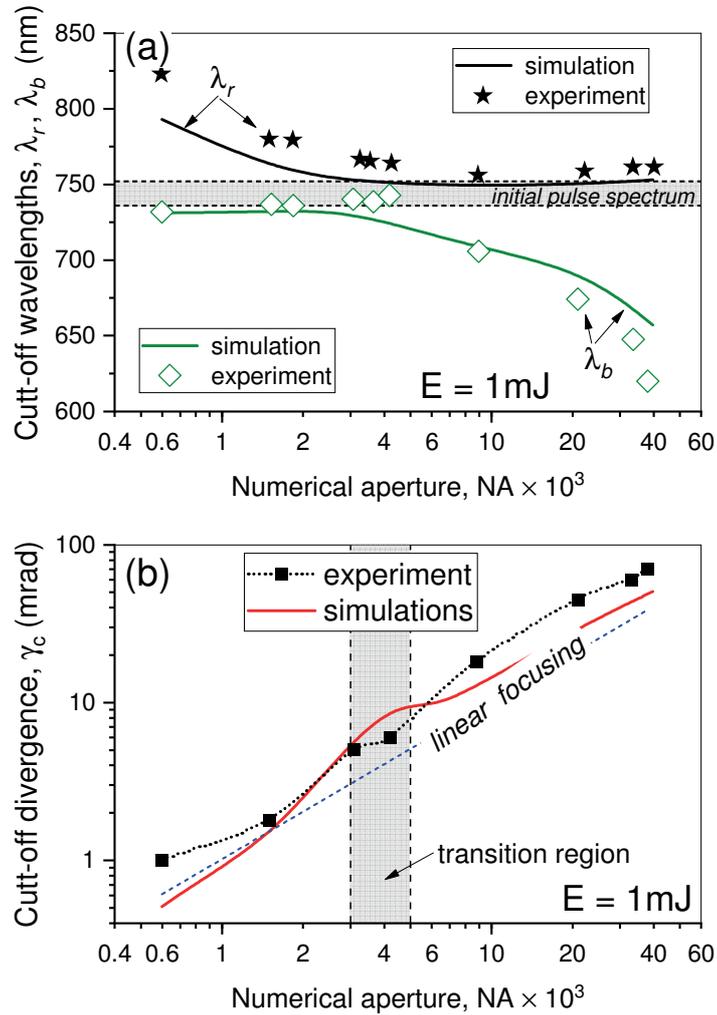

Fig. 3. (a) Maximum $\lambda_r$ (circles) and minimum $\lambda_b$ (squares) cut-off wavelengths of the laser pulse spectrum and (b) cut-off spatial angle $\gamma_c$ after the filamentation in air as a function of a numerical aperture *NA*. The initial laser pulse spectrum and linear focusing transition region are marked by shaded areas.

In the intermediate group combining the frequency-angular spectra with the numerical apertures from $NA = 3 \cdot 10^{-3}$ to $5 \cdot 10^{-3}$ (Figs. 2(e)-(g)), both spectral and angular broadening after the filamentation is weakly pronounced. According to [12], we can claim that these spectral distributions relate to the region where a transition from linear to nonlinear (filamentation) pulse focusing takes place. In this case, as can be seen in Fig. 1(d) for $NA = 3 \cdot 10^{-3}$, the filament length and plasma channel density are low and cannot provide a sufficient level of nonlinear pulse spectral transformations under propagation in the medium. In this case the total level of the laser pulse nonlinear phase increment accumulated in the interaction region (B-integral [9]) is not high enough for emergence of a noticeable maximum in the long-wavelength region. Meanwhile, due to relatively low plasma density, the pulse spectrum broadening to the shorter wavelengths is also minimal.

Worthwhile noting, in Figs. 2(b), (f) and (i) the spectra obtained for a laser pulse with the initial diameter $D_0 = 8$ mm are shown, which is almost three times larger than one in other figures.

Nevertheless, the trend of changing the spectral-angular characteristics in dependence on the numerical aperture just persists. This proves the fact, that the dimensionless parameter of numerical aperture influences the pulse spectrum transformations during the laser pulse propagation under filamentation in focusing regime.

In Figs. 3(a) and (b) the cut-off wavelengths in blue $\lambda_b$ and red $\lambda_r$ spectral regions as well as cut-off spatial angle $\gamma_c$ taken at the level of 1/10 from the spectral intensity maximum - after the focused laser pulse filamentation with initial energy $E = 1$ mJ - are presented. Taking into account the spectral-angular distributions in Fig. 2 one can see, that under the tight focusing conditions the laser pulse spectrum is broadened mainly to the anti-Stokes spectral region, and $\lambda_b$ increases. On the contrary, in the case of weak focusing, the pronounced spectral red shift is observed causing the growth of $\lambda_r$. In the transitional range of $NA$ from $3 \cdot 10^{-3}$ to $5 \cdot 10^{-3}$ the laser pulse after the filamentation exhibits a minimal spectral broadening.

Generally, the cut-off divergence angle $\gamma_c$ demonstrates $NA$-dependence similar to the linear focusing regime, i.e. $\gamma_c \sim NA$. The detailed analysis of this dependence shows that there are three characteristic regions, where the second derivative $\partial^2 \gamma_c / \partial (NA)^2$ is positive, negative or changes its sign. These regions correspond, respectively, to the domination of nonlinear pulse self-focusing over the linear diffraction for low $NA$, domination of the geometric focusing for high $NA$-values, and close to neutral divergence angle behavior under the conditions when the partial impacts of the Kerr-nonlinearity and linear focusing are equal. The latter $NA$-region can be addressed to the linear-to nonlinear focusing transition.

## 3. Energy limit for linear pulse focusing:numerical simulations

For various practical applications it is important to know -for given $NA$ - the limiting laser pulse power and energy, at which the laser beam focusing in a nonlinear medium still resembles a linear focused propagation, i.e. the focal waist dimensions and its spatial position can be determined according to linear diffraction laws. In other words, one need to find the energy threshold for focused laser pulse propagation under conditions of the Kerr self-focusing and plasma formation in the medium, when the laser beam focusing occurs without significant nonlinear transformations of the pulse transverse profile and its spectral composition, i.e. to find the above-mentioned transition region between the linear-to-nonlinear laser pulse focusing.

To determine the linear focusing threshold in a nonlinear medium, we perform the numerical simulations of a pre-focused femtosecond laser pulse filamentation in air. In the numerical experiments, the scalar unidirectional pulse propagation equation (UPPE) is solved with a carrier resolution by accounting the optical nonlinearity of air as is detailed in [20, 21]. The model of air optical nonlinearity includes instant and retarded components of the Kerr effect, high order nonlinearities (saturation of the Kerr nonlinearity), complex refraction index changes due to photoionization of air molecules and formation of free-electrons gas. The linear part of the UPPE accounts for group velocity dispersion of femtosecond laser pulse in air (according to Cauchy formula) and diffraction by taking into account the non-paraxial nature of the focused beam propagation. The dynamic of the volumetric free-electrons concentration $\rho_e$ of plasma channel emerging during pulse propagation is calculated using corresponding rate equation considering the combined multiphoton/tunneling mechanism of air gas components (nitrogen, oxygen) ionization following the Perelomov-Popov-Terent'ev (PPT) model and also recombination losses of electron concentration [22]. The problem is solved in axisymmetric (2D+1) formulation for the initial laser pulse parameters corresponding to the experimental values. Normally, the calculation domain is discretized by approximately $2^{24}$ mesh nodes and the task is launched on 72 cores of the NUSC supercomputer cluster [23] based on Intel Xeon E5-2680v3 blade servers. The average program runtime is from 40 to 100 hours while using adaptive step on evolutional variable and varying *NA* from $50 \cdot 10^{-3}$ to $0.5 \cdot 10^{-3}$.

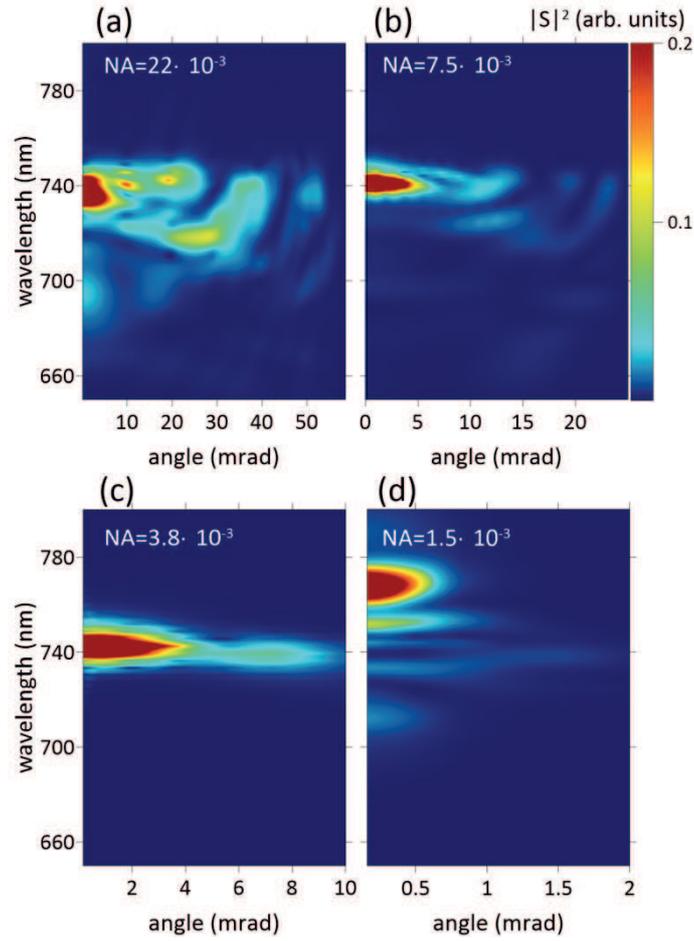

Fig. 4. Numerically simulated frequency-angular distribution of optical intensity $|S|^2$ for the laser pulse with initial energy $E = 1$ mJ under various focusing conditions.

The results of our numerical simulations of a laser pulse focused propagation with $D_0 = 3$ mm and initial energy $E = 1$ mJ in air are shown in Fig. 1(e) as the longitudinal profiles of the linear free-electrons density $\rho_{ez}$. This parameter is calculated as the integral of volumetric plasma density $\rho_e$ over the cross-section of the calculation domain. In Fig. 3(a) the cut-off wavelengths of pulse spectrum, $\lambda_b$ and $\lambda_r$, are presented, respectively, in anti-Stokes and Stokes spectral regions. Similarly, the calculated cut-off pulse divergence angle $\gamma_c$ is shown in Fig. 3(b). These parameters are calculated at the level of the spectral intensity drop by 10 dB in the spectra obtained at the distance of four focal lengths.

The frequency-angular distribution of pulse intensity $|S(\lambda,\gamma)|^2$ calculated for several numerical apertures *NA* is shown in Figs. 4(a)-(d). The thorough comparison of the simulated and experimental dependences shown in the aforementioned figures indicates their good qualitative agreement for all the considered pulse focusing conditions.

To determine the threshold energy of a focused laser pulse, when the linear focusing regime transits to the nonlinear one, we carried out a corresponding parametric study, the results of which are shown in Figs. 5(a) and (b). Thus, Fig. 5(a) represents the simulated spectra of a laser pulse (the initial energy $E = 1$ mJ) after its nonlinear propagation over four focal distances corresponding to the three numerical apertures $NA = 0.6 \cdot 10^{-3}$, $3 \cdot 10^{-3}$ and $20 \cdot 10^{-3}$. For every case, the laser pulse spectral width $\Delta\lambda$ is taken from the corresponding spectrum at the level of tenfold spectral radiance drop from its maximal value as shown by the horizontal dashed lines in the figure.

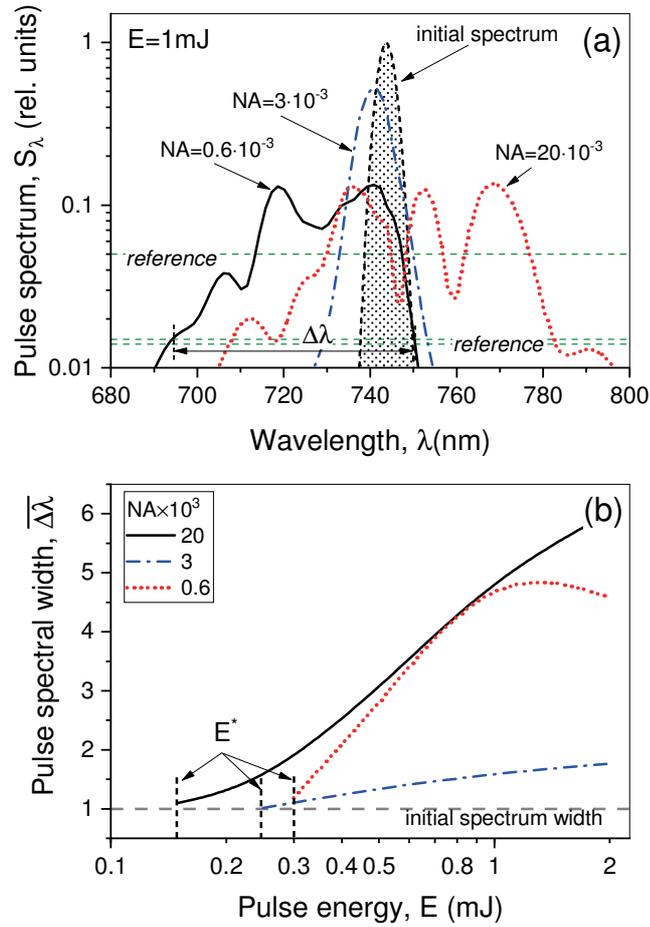

Fig. 5. (a) Simulated laser pulse spectra ($E$= 1 mJ) after the filamentation for different numerical apertures *NA*. Dashed horizontal lines show the levels at which the spectral width $\Delta\lambda$ is measured. The initial pulse spectrum is represented by the shadowed area; (b) Relative spectral width $\overline{\Delta\lambda}$ versus pulse energy $E$ and numerical aperture *NA*. Vertical lines depict the threshold energy $E^*$ demarcating the regimes of linear and nonlinear focusing for each *NA*.

These spectral distributions distinctively show qualitative difference of spectral transformations depending on initial geometrical pulse focusing. On the one hand, at any *NA* we observe a retaining trend for the laser pulse spectrum to expand toward the blue (anti-Stokes) region under pulse energy growth. This reflects the increase of plasma nonlinearity impact on the pulse self-phase modulations. On the other hand, under the tight and moderate spatial focusing, the long-wavelength wing of the pulse spectrum is almost independent of the initial pulse energy and its limiting value $\lambda_r$ remains almost unchanged comparing to the initial one. However, in the case of weak initial focusing ($NA = 0.6 \cdot 10^{-3}$) the spectral broadening to the red spectral region due to rotational Raman scattering on the nitrogen molecules becomes quite noticeable and prevails over the blue spectral shift.

The dependence of relative spectral width $\overline{\Delta\lambda} = \Delta\lambda/\Delta\lambda_0$, where $\Delta\lambda_0$= 12 nm is the initial laser pulse spectral width, on the pulse energy $E$ for different numerical apertures is shown in Fig. 5(b). The threshold pulse energy $E^*$ for linear focusing is defined as the energy, at which pulse spectral width $\Delta\lambda$ measured after the geometrical focus is 10% higher than the initial value $\Delta\lambda_0$. In Fig. 5(b) this limit is indicated by the vertical dashed lines.

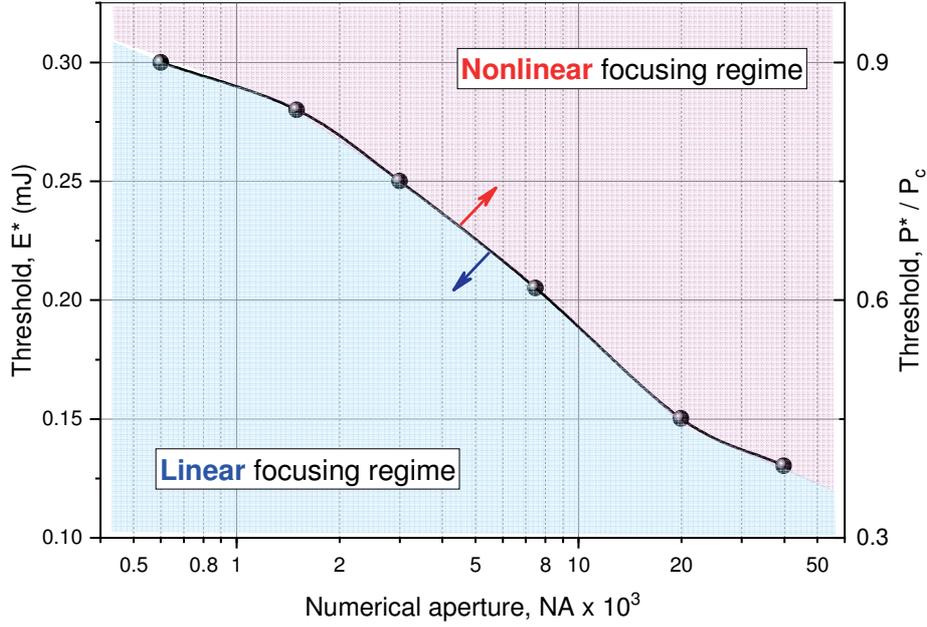

Fig.6. Threshold energy $E^*$ and threshold peak power $P^*$ of the laser pulse linear focusing at different numerical apertures *NA*.

The dependence of $E^*$ on the laser pulse *NA* is presented in Fig. 6. On the right vertical axis of the diagram we plot additionally the scale for threshold peak power $P^*$ calculated from $E^*$ in accordance with the experimental pulse parameters. It is quite interesting that under the tight focusing the limit, at which the optical nonlinearity of propagation medium manifests itself, lies well below the critical power $P_c$ for the pulse self-focusing. This can be explained by the fact that the theoretical parameter $P_c$ was historically deduced from the analysis of a collimated laser beam self-focusing dynamics, so, it does not account for cumulative-like medium ionization under the laser beam transverse collapsing in the vicinity of the nonlinear focus. The tighter is the pulse focusing, the more $P^*$ differs from $P_c$. Meanwhile, for a collimated beam under the limit $NA \to 0$, the threshold laser pulse power approaches the critical one, i.e. $P^* \to P_c$.

From a practical point of view, the spatial structure of a laser pulse propagating in regime of filamentation under initial geometrical focusing is of considerable interest. Furthermore, as stated above, it is important to achieve maximum pulse intensity in the focal spot providing that possible nonlinear distortions of a transversal laser beam profile remains insignificant, whereas the actual focal point spatial position coincides with one calculated in linear diffraction limit.

Numerically simulated transverse dimensions of a femtosecond laser pulse focused in ambient air under different *NA* and *E* are shown in Figs. 7(a) and (b). Here we present the graph of relative beam transverse area $\overline{A}(z) = [R(z)/R_L(z)]^2$, which is derived from the dynamics of pulse transverse radius $R(z)$. We plot this parameter in the dependence on both absolute focal shift $\Delta z = z-f$ and relative shift $\overline{\Delta z} = \Delta z/z_R$, i.e. divided by the focal waist length $z_R = 1/[2k(NA)^2]$, where $k = 2\pi/\lambda$. The laser pulse radius $R$ in its turn is defined at $1/e$-level of the maximal fluence. Parameter $R_L$ in the denominator of $\overline{A}$ accounts for laser beam radius during focused propagation in a linear medium ($\eta \ll 1$) and is obtained from the well-known formula for a Gaussian beam: $R_L(z) = R_0\sqrt{(z/2L_R)^2 + (1-z/f)^2}$, where $R_0 = R(z=0)$, and $L_R = kR_0^2/2$ is the Rayleigh length forthe initial laser beam. Thus, any deviation of relative pulse area $\overline{A}(z)$ from unity evidences the transition from the linear focusing regime.

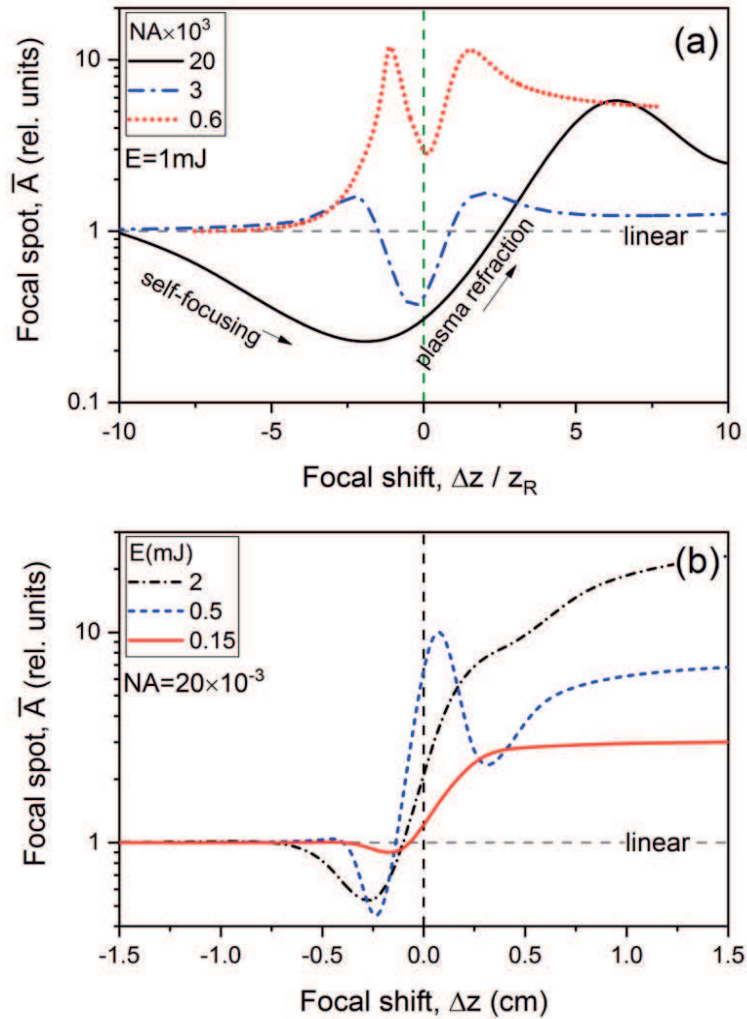

Fig. 7. Relative laser beam focal waist area $\overline{A}$ versus propagation distance at different (a) numerical apertures $NA$ and (b) pulse energy $E$.

By analyzing the dependencies given in Fig. 7(a) we can see that the laser pulse focusing with the supercritical peak power ($\eta \approx 3$) takes place in the nonlinear regime at any considered numerical aperture. The largest pulse spatial area deviations from the linear focusing regime occur near the geometrical focal point ($\Delta z = 0$) at the weakest $NA = 0.6 \cdot 10^{-3}$ or, on the contrary, under the tightest focusing ($NA = 20 \cdot 10^{-3}$). It should be noted, that in the latter case from the dependence $\overline{A}(\Delta z)$ one can trace both the influence of the Kerr self-focusing, which is clearly seen as the values $\overline{A} < 1$ before the geometrical focus, and plasma refraction after the focal point giving $\overline{A} > 1$. Minimal nonlinear beam size distortions are observed at moderate numerical apertures, when spectral-angular transformations of a laser pulse are less pronounced also (see Figs. 2 and 3).

It is worth to pay attention to the pulse nonlinear spatial distortions lowering in the focal area with the decrease in laser pulse power. As can be seen in Fig. 7(b) for the case with a tightly focused pulse with $NA = 20 \cdot 10^{-3}$, the pulse energy decreasing down to $E = 0.15$ mJ, that corresponds to the limit energy $E^*$ (Fig. 6), almost completely eliminates the influence of air optical nonlinearity on the spatial beam profile. Actually, the changes of pulse cross-sectional area before the geometrical focus does not exceed 10%, and the nonlinear focus $f_N$ itself is the closest to the geometrical focus ($f = 7.5$ cm) with the longitudinal shift $\Delta z$ of only $\approx$ -1 mm. At the same time, even at the energy of $E = 0.5$ mJ one can observe more than twofold drop of pulse transverse area in the nonlinear focus and multiple increase of the focal waist shift up to $\Delta z \approx$ -2.5 mm.

## 4. Conclusions

In conclusion, we consider the problem of a femtosecond laser pulse nonlinear propagation in air under various focusing conditions for the assessment of spectral-angular and spatial laser pulse transformations when entering the filamentation regime. By changing the initial numerical aperture of the laser pulse we experimentally obtain and numerically simulate the frequency-angular spectra and linear plasma density profiles in the focal area. We show that under the tight spatial focusing ($NA > 5 \cdot 10^{-3}$), the laser pulse with the supercritical power undergoes spectral broadening mainly toward the short-wavelength region and after the filamentation propagates at highest angles to the optical axis. Under the weak focusing conditions ($NA < 2 \cdot 10^{-3}$), the pulse spectrum broadens mostly to the Stokes spectral wing, whereas the blue-shift becomes diminished, and the pulse spatial divergence after the nonlinear focus remains low. Our experiments and numerical simulations indicate the existence of the specific range of pulse numerical apertures approximately from $NA = 2 \cdot 10^{-3}$ to $5 \cdot 10^{-3}$, within which the broadening of pulse frequency-angular spectra during nonlinear focusing practically does not occur. This $NA$-range can be conditionally attributed to the previously reported in [12] transition region from linear- to nonlinear pulse focusing.

By means of the computer simulations of focused femtosecond laser pulse filamentation in air we derive the energy threshold $E^*$ as the limiting factor between linear (diffraction) and nonlinear (filamentation) focusing regimes in a wide range of numerical apertures. In our study we take into account the influence of both the optical Kerr effect and plasma nonlinearity. We demonstrate that for a given $NA$, the laser pulse energy excess over this threshold level $E^*$ not only leads to the frequency spectrum broadening but also noticeably affects the spatial structure of the laser beam in the vicinity of the focal point. Specifically, an increase in the pulse angular divergence and large-scale spatial distortions of the focusing spot are observed. At the same time, fulfilling the condition $E_0 \approx E^*$ does guarantee the maximal pulse intensity in the focus while maintaining good quality of spectral and spatial pulse shape.

Our findings clearly define the range of laser beam numerical apertures and pulse energies providing maximal laser intensities with good beam quality in the focal point that can be crucial for various laser microstructuring and micromachining technologies.

**Funding**

The research was partially supported by the Ministry of Science and Higher Education of the Russian Federation (V.E. Zuev Institute of Atmospheric Optics of Siberian Branch of the Russian Academy of Sciences); Russian Science Foundation (Agreement #21-12-00109); Russian Foundation for Basic Research (Grant #20-02-00114).

# References

1. Couairon, A. and A. Mysyrowicz. "Femtosecond filamentation in transparent media." Physics reports 441.2-4 (2007): 47-189.
2. Liang, H. et al. "Three-octave-spanning supercontinuum generation and sub-two-cycle self-compression of mid-infrared filaments in dielectrics." Optics letters 40.6 (2015): 1069-1072.
3. Kosareva, O. G., et al. "Conical emission from laser–plasma interactions in the filamentation of powerful ultrashort laser pulses in air." Optics letters 22.17 (1997): 1332-1334.
4. Chin, S. L., et al. "Advances in intense femtosecond laser filamentation in air." Laser Physics 22.1 (2012): 1-53.
5. Chekalin, S. V., and V. P. Kandidov. "From self-focusing light beams to femtosecond laser pulse filamentation." Physics-Uspekhi 56.2 (2013): 123.
6. Marburger, J. H. "Self-focusing: theory." Progress in quantum electronics 4 (1975): 35-110.
7. Kiselev, D., Ludger W., and J-P. Wolf. "Filament-induced laser machining (FILM)." Applied Physics B 100.3 (2010): 515-520.
8. Ionin, A. A., et al. ""Heterogeneous" versus "homogeneous" nucleation and growth of microcones on titanium surface under UV femtosecond-laser irradiation." Applied Physics A 116.3 (2014): 1133-1139.
9. Talanov, V. I. "Focusing of light in cubic media." ZhETF Pisma Redaktsiiu 11 (1970): 303.
10. Théberge, F. et al. "Plasma density inside a femtosecond laser filament in air: Strong dependence on external focusing." Physical Review E 74.3 (2006): 036406.
11. Geints, Yu E., et al. "Peculiarities of filamentation of sharply focused ultrashort laser pulses in air." Journal of Experimental and Theoretical Physics 111.5 (2010): 724-730.
12. Lim, K. et al. "Transition from linear-to nonlinear-focusing regime in filamentation." Scientific reports 4.1 (2014): 1-8.
13. Reyes, D. et al. "Transition from linear-to nonlinear-focusing regime of laser filament plasma dynamics." Journal of Applied Physics 124.5 (2018): 053103.
14. Dergachev, A. A. et al. "Filamentation of IR and UV femtosecond pulses upon focusing in air." Quantum Electronics 43.1 (2013): 29.
15. Geints, Yu E., et al. "Kerr-driven nonlinear refractive index of air at 800 and 400 nm measured through femtosecond laser pulse filamentation." Applied Physics Letters 99.18 (2011): 181114.
16. Geints, Yu E., et al. "High intensive light channel formation in the post-filamentation region of ultrashort laser pulses in air." Journal of Optics 18.9 (2016): 095503.
17. Ionin, A. A., et al. "Multiple filamentation of intense femtosecond laser pulses in air." JETP letters 90.6 (2009): 423-427.
18. Kosareva, O. et al. "Tunable infrared supercontinuum on 100 m path in air." Optics Letters 46.5 (2021): 1125.
19. Chen, Y., et al. "Observation of filamentation-induced continuous self-frequency down shift in air." Applied Physics B 91.2 (2008): 219-222.
20. Couairon, A., et al. "Practitioner's guide to laser pulse propagation models and simulation." The European Physical Journal Special Topics 199.1 (2011): 5-76.
21. Geints, Yu E., and A. A. Zemlyanov. "Ring-Gaussian laser pulse filamentation in a self-induced diffraction waveguide." Journal of Optics 19.10 (2017): 105502.
22. Perelomov, A. M., V. S. Popov, and M. V. Terent'ev. "Ionization of atoms in an alternating electric field." Sov. Phys. JETP 23.5 (1966): 924-934.
23. Internet link: http://nusc.nsu.ru/wiki/doku.php/doc/index